\documentstyle[12pt]{article}

\begin{document}
 \title{Empirical logic of finite automata: microstatements versus
macrostatements}
\author{K. Svozil\thanks{Institut f\"ur Theoretische Physik,
   University of Technology Vienna,  Wiedner Hauptstra\ss e 8-10/136,
 A-1040 Vienna, Austria; e-mail:  svozil@tph.tuwien.ac.at}
 \hspace*{2mm}  and
R. R. Zapatrin\thanks{Department of Mathematics, SPb UEF,
Griboyedova, 30-32, 191023,
St. Petersburg, Russia; e-mail: tmpry@friedman.usr.lgu.spb.su}
}

\maketitle

\begin{abstract}
We compare the two approaches to the empirical logic of automata.
The first, called partition logic (logic of microstatements), refers to
experiments on individual
automata. The second one, the logic of simulation (logic of
macrostatements), deals with ensembles of automata.
\end{abstract}

\section*{Introduction}

The principal goal of this paper is to bring the two
approaches to the empirical logic of automata in a coherent
perspective. Two groups of researchers (M.~Schaller and K.~Svozil
in Vienna, and A.~A.~Grib and R.~R.~Zapatrin in St.~Petersburg) have
approached this issue from different directions. In this joint paper we
investigate the connection between these two approaches.

The paper is organized in the following way: in section \ref{s1} the
class of automata we are going to deal with is specified. In section
\ref{s2} the concept of the empirical logic of automata is
specified. The approach carried out by the Vienna group based on a
partitioning of the set of automaton states is introduced in section
\ref{s3}, and that of the St.~Petersburg group based on a closure
operation on the set of inputs is described in section \ref{s4}. The
comparison of the two approaches concludes the paper.

\section{Automaton model}
\label{s1}
In this paper we shall deal with the specific class of automata, called
{\sc normalized} introduced in \cite{grzap90}. The main feature of these
automata is that they may be completely defined by a non-oriented graph
$P=(V,E)$, $V$ being the set of vertices and $E$ the set of edges. The
states of the automaton correspond
to the vertices of the graph $P$. We also assume the existence of
a so-called final state $0$. The inputs of the automaton are in
one--to--one
correspondence with its states. One could think of applying an input
$v$ as checking the appropriate state. Now assume that the automaton is
initially in state $u$ ($\neq 0$) and input $v$ is
applied. Then the {\sc transition} $\delta :u\rightarrow v$
takes place if and only if the vertices $u,v$ are linked by an edge of
the graph $P$, otherwise the automaton goes into the final state $0$;
i.e.,
\[ \delta (u,v) := \left\lbrace
\begin{array}{rcl}
v&,&\mbox{ if }(u,v)\in E\\
0&&\mbox{ otherwise}
\end{array}
\right.
\]
The two-valued {\sc output function} depends on the
{\em resulting}
state. Its value is $0$ if the resulting state is the final state, and
$1$ if the resulting state is any other; i.e.,
\[ \lambda (\delta (u,v)) := \left\lbrace
\begin{array}{rcl}
1&,&\mbox{ if }\delta (u,v)\neq 0\\
0&,&\mbox{ if }\delta (u,v)= 0
\end{array}
\right.
\]

\begin{figure}[hb]
\unitlength 1.00mm
\begin{center}
\begin{picture}(24,20)
\put(20,5){\line(0,1){15}}
\put(5,5){\line(1,0){15}}
\put(5,20){\line(1,0){15}}
\put(5,5){\circle*{2}}
\put(20,5){\circle*{2}}
\put(20,20){\circle*{2}}
\put(5,20){\circle*{2}}
\put(0,20){1}
\put(22,20){2}
\put(22,0){3}
\put(0,0){4}
\end{picture}
\end{center}
\caption{The graph $P(V,E)$. The set of vertices $V=\{1,2,3,4\}$, the
set of edges $E=\{(1,2),(2,3),(3,4)\}$.}
\label{f124}
\end{figure}
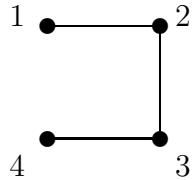

As an example, consider the normalized automaton associated with the
graph $P$ drawn in Fig. \ref{f124}.
Suppose the automaton is prepared in the initial state corresponding to
the vertex $1$. Then, if we apply the input $1$, the state of the
automaton will not change, and it will output the symbol $1$. If,
starting from the same initial state $1$, the input $2$ is applied,
the state of the automaton changes to $2$, though the output value will
still be $1$. If, on the other hand, we apply any one of the inputs $3$
or
$4$, then the automaton makes a transition to the final state $0$
(since there is no edge between the initial vertex $1$ and the vertices
$3$ and $4$, respectively) and the output value will be $0$.
We shall return to this example in the discussion of the propositional
structures below.

\section{Empirical logic of automata}
\label{s2}

The motivation behind the investigations in this paper is the
construction of primitive  {\em empirical statements} or {\em
propositions} about automata \cite{finkelstein}.
Such  experimental statements
form the basis of the formal investigation of the corresponding  logics.
In particular, there exist automata for which
validation of one empirical statement
makes impossible the validation of another empirical statement and
{\it vice versa}, as it was first pointed out by Moore \cite{moore}.

Thereby, one decisive feature of the set-up is the {\em intrinsic}
character of the measurement process:
The automaton is treated as a black box with known
description but unknown initial state.
Automata experiments are conducted by applying an input sequence and
observing the output sequence.

In the above example, it is not necessary to input sequences containing
more than one symbol, since the non-final states are not distinguished
by the output function.

\section{Microstatements: partition logic}
\label{s3}

The conventional state identification problem \cite{moore,conway,bauer}
is to obtain
information about an unknown initial state.  Thereby it is assumed that
only {\em a single} automaton copy is available for inspection.  That
is, no second, identical example of the automaton can be used for
further examination.  Alternatively, one may think of it as choosing at
random a single automaton from an automata ensemble which differ only by
their initial state.  The task then is to find out which was the initial
state of the chosen automaton.

The partition logic approach to finite automata study was
introduced in
\cite{svozil-93}, and subsequently in \cite{svosh,svosh2,svosh3}.
There,
statements about single automata of the following
form $p_A$ are
considered:
\begin{equation}
p_A:=\mbox{"the state $v$ of the automaton is in $A$"},
\label{pa}
\end{equation}
where $A$ is a subset of the set of automaton states.
For any statement $p_A$ we can build its opposite $\overline{p_A}$,
namely,
\[p_A:=\mbox{"the state $v$ of the automaton is NOT in $A$"}\]

Let us try to characterize this state identification problem
algebraically. Again, $V$ denotes the set of automaton states.
Now associate with any input $v$ the
set of automaton states
such that any pair of states from any element of the partition are
mutually indistinguishable.
For normalized automata studied in this paper, this partition reduces to
the splitting of the entire set of states $V$ into two subsets
\begin{eqnarray}
V&=& V_0 \cup V_1\label{part-1}\\
V_0&=&\{w\in V\mid (v,w)\not\in E \}\\
V_1&=&\{w\in V\mid (v,w)\in E \}
\end{eqnarray}
where $E$ stands for the set of edges of the graph $P(V,E)$.
Then, a property $p_A$ of the form (\ref{pa}) is said to be
{\sc testable} if and only if the set $A$ is contained in the partition
(\ref{part-1}) associated with some input.

Now consider the entire collection of the sets of the form $V_0$ and
$V_1$ (\ref{part-1}) for all possible inputs $v$.
Note that for different inputs, the partitions (\ref{part-1}) may
coincide (cf. the example below).  $V_0$ may be thought of as the
negation of $V_1$ and vice versa. Therefore, each pair of them
form a Boolean algebra $2^2$. The propositional structure is then built
from these algebras by pasting; i.e., by identifying their greatest and
their least elements. The resulting propositional
structures are the lattices
$MOn$, where $n$ stands for the number of different partitions
(\ref{part-1}).
The partial order can be interpreted as logical implication
``$\Rightarrow$'' on the statements about a {\em single} automaton from
the ensemble. This is why we call these statements {\sc
microstatements}.

\begin{figure}
 \begin{center}
\unitlength 1.00mm
\linethickness{0.4pt}
\begin{picture}(113.28,55.00)
\put(62.33,10.00){\circle*{1.89}}
\put(62.33,50.00){\circle*{1.89}}
\put(52.33,30.00){\circle*{1.89}}
\put(32.66,30.00){\circle*{1.89}}
\put(12.66,30.00){\circle*{1.89}}
\put(112.33,30.00){\circle*{1.89}}
\put(92.66,30.00){\circle*{1.89}}
\put(72.66,30.00){\circle*{1.89}}
\multiput(62.33,50.00)(0.30,-0.12){167}{\line(1,0){0.30}}
\multiput(112.33,30.00)(-0.30,-0.12){167}{\line(-1,0){0.30}}
\multiput(62.33,50.00)(0.18,-0.12){167}{\line(1,0){0.18}}
\multiput(92.33,30.00)(-0.18,-0.12){167}{\line(-1,0){0.18}}
\multiput(62.33,10.00)(0.12,0.23){84}{\line(0,1){0.23}}
\multiput(72.33,29.67)(-0.12,0.24){84}{\line(0,1){0.24}}
\multiput(62.33,50.00)(-0.12,-0.23){84}{\line(0,-1){0.23}}
\multiput(52.66,29.67)(0.12,-0.24){81}{\line(0,-1){0.24}}
\multiput(62.33,10.00)(-0.18,0.12){164}{\line(-1,0){0.18}}
\multiput(32.66,29.67)(0.17,0.12){170}{\line(1,0){0.17}}
\multiput(62.33,50.00)(-0.30,-0.12){167}{\line(-1,0){0.30}}
\multiput(12.33,30.00)(0.30,-0.12){167}{\line(1,0){0.30}}
\put(62.33,5.00){\makebox(0,0)[cc]{$\emptyset$}}
\put(62.33,55.00){\makebox(0,0)[cc]{$\{1,2,3,4\}$}}
\put(2.00,30.00){\makebox(0,0)[lc]{$ \{1,2\}$}}
\put(23.00,30.00){\makebox(0,0)[lc]{$\{3,4\}$}}
\put(39,30.00){\makebox(0,0)[lc]{$\{1, 2,3\}$}}
\put(62.66,30.00){\makebox(0,0)[lc]{$\{4\}$}}
\put(82.66,30.00){\makebox(0,0)[lc]{$\{1\}$}}
\put(114.00,30.00){\makebox(0,0)[lc]{$\{2,3,4\}$}}
\end{picture}
 \end{center}
 \caption{Lattice $MO3$ of the intrinsic propositional calculus.}
\label{f-mo3}
\end{figure}
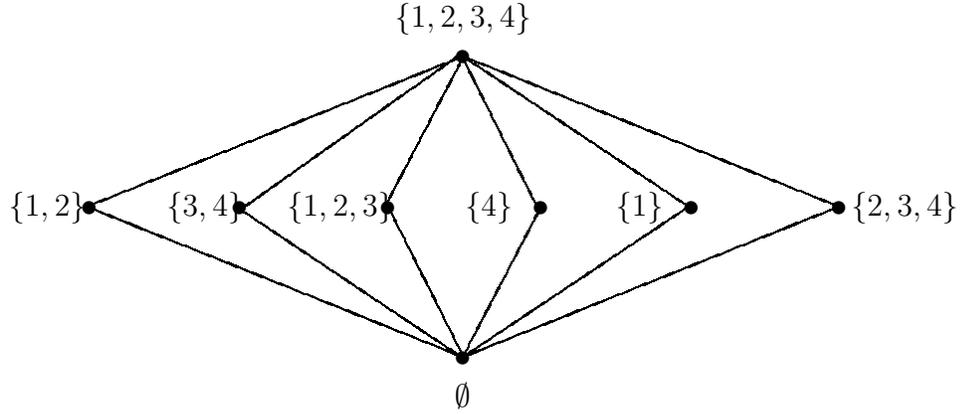

To take up the example mentioned earlier, consider the partitions
produced by all possible inputs $v=1,2,3,4$. They are
\begin{eqnarray}
v=1&:&V=\{1,2\} \cup \{3,4\}\\
v=2&:&V=\{1,2,3\} \cup \{4\}\\
v=3&:&V=\{1\} \cup \{2,3,4\}\\
v=4&:&V=\{1,2\} \cup \{3,4\}
\end{eqnarray}
Note that the inputs $1$ and $4$ produce the same partitions.
The resulting propositional structure is drawn in Fig. \ref{f-mo3}.
It should be emphasized that although the elements of the lattice
associated with the partition logic are the subsets of the set of states
$V$, the partial order relation (``implication
$\Rightarrow$'')
is weaker than the set theoretical inclusion ``$\subset$.'' For
instance, we can see from the above example that
$\{1\} \subset
\{1,2,3\}$
but
$\{1\} \not\Rightarrow
\{1,2,3\}$. In general, the partial order relation always implies set
inclusion but not vice versa. Only if two statements can be
identified by the same experiment, i.e., only if both statements are
simultaneously measurable, the partial ordering coincides with set
inclusion.

\section{Macrostatements: logic of simulation}
\label{s4}
Another approach to the logic of finite automata was introduced in
\cite{grzap90,grzap92}. Unlike that giving rise to the partition logic
described
in the previous section, it assumes dealing with ensembles of
automata
rather than with a single pattern and therefore the notion of property
is referred to the {\em ensemble}. This was the reason to call the
statement arising in this approach {\sc macrostatements}.

Before describing the approach in general, we return to our example of
automaton. Suppose now that we have at our disposal two ensembles of
such automata.
The first of them is prepared in such a way that some of
the constituent automata of the ensemble are in the state
$\{1,2,3\}$.
That is, initially, the automata from the first ensemble are
either in state $1$ or in state $2$ or in state $3$.
The second ensemble is prepared
in the state
$\{2,3,4\}$.
That is, initially, the automata from the first ensemble are
either in state $2$ or in state $3$ or in state $4$.

As before,
only one experiment on each particular automaton from the ensemble can
be carried out; from each individual automaton in the
ensemble, only a single copy is available.

We first perform experiments on the first ensemble.
Suppose we have generated the protocol of a sufficiently large
sequence
of measurements of the results of each input. The values of the output
function which can be observed are
$0$ and $1$ for each input.

Next, we deal with the second ensemble in a similar way.
Again, looking into the protocol, we see that there are $0$'s and $1$'s
as the result of each input. Since we have made no particular choice of
the distribution of the initial states, we have to conclude that the two
states
of the ensemble described above are indistinguishable.

However, assume that
a third ensemble is prepared in the "pure" state associated with the
activation
of only the vertex $1$. Then, analyzing the protocol, one finds that the
output values corresponding to the input $3$ are always $0$.
This makes the protocol differ from the protocols generated from the
first two ensembles. Thus, we can distinguish the state of the third
ensemble from the states of the first and the second ensemble.

We now give the rigorous definitions. With each subset $A\subseteq V$,
we
consider the statement about the ensemble of automata of the following
form $s_A$:
\begin{equation}
s_A:=\mbox{"the state $v$ of each particular automaton is in $A$"}
\label{sa}
\end{equation}
For any statement $s_A$ we can build its opposite $\overline{s_A}$,
namely,
\[s_A:=\mbox{"the state $v$ of each particular automaton is NOT in
$A$"}\]
Then, a property $s_B$ of the form (\ref{sa}) is said to be
{\sc testable} if and only if there exist a subset $A\subseteq V$
such that \[ s_B=\overline{s_A} \]

The motivation for this definition is the following. The outcome of
the statistical experiment on the ensemble of automata is the protocol
of the following form:
\[ \begin{array}{ll}
\mbox{input 1}&0001110101\ldots\\
\mbox{input 2}&1001010111\ldots\\
\ldots&\ldots\\
\vdots&\vdots\\
\mbox{an input}&\mbox{the values of the output function}
\end{array}
\]
The rows of the protocol containing the output values equal to $1$
bring no essential information.
Only
the rows in which  the values of the output
function is always $0$ contain information relevant to our purposes.
That is the reason why
the subsets having only $0$'s in the appropriate line of the protocol
are treated as testable within this approach.

\begin{figure}[hb]
\begin{center}
\unitlength 1mm
\linethickness{0.4pt}
\begin{picture}(50.00,70.00)
\put(30.00,10.00){\circle*{2.11}}
\put(45.00,25.00){\circle*{2.11}}
\put(45.00,45.00){\circle*{2.11}}
\put(15.00,25.00){\circle*{2.11}}
\put(15.00,45.00){\circle*{2.11}}
\put(30.00,65.00){\circle*{2.11}}
\put(30.00,10.00){\line(-1,1){15.00}}
\put(15.00,25.00){\line(0,1){20.00}}
\put(15.00,45.00){\line(3,4){15.00}}
\put(30.00,65.00){\line(3,-4){15.00}}
\put(45.00,45.00){\line(0,-1){20.00}}
\put(45.00,25.00){\line(-1,-1){15.00}}
\put(29.67,5.00){\makebox(0,0)[cc]{$\emptyset$}}
\put(30.00,70.00){\makebox(0,0)[cc]{$V$}}
\put(10.00,45.00){\makebox(0,0)[rc]{$\{1,2\}$}}
\put(50.00,45.00){\makebox(0,0)[lc]{$\{3,4\}$}}
\put(50.00,25.00){\makebox(0,0)[lc]{$\{4\}$}}
\put(10.00,25.00){\makebox(0,0)[rc]{$\{1\}$}}
\end{picture}
\end{center}
\caption{The diagram of testable properties.}
\label{figm4}
\end{figure}
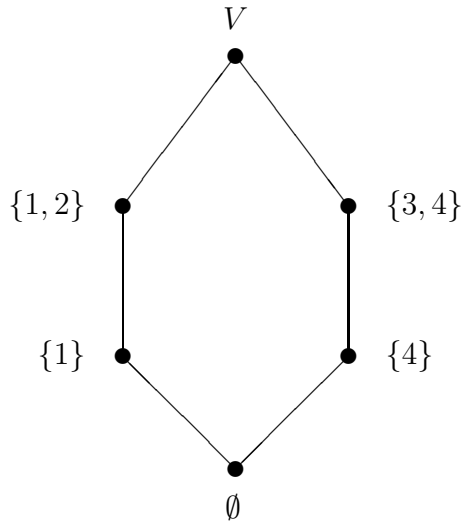

In our example, the only proper (i.e. $\neq 0,V$)
subsets which are in principle testable, are:
$\{1\},\{4\},\{1,2\},\{3,4\}$. For instance, if, on any input of $1$,
the output is always $0$, then one can conclude that the automata in the
ensemble are either in the (micro-)state $3$ or in (micro-)state $4$;
therefore the ensemble is in the macrostate denoted by $\{3,4\}$.
Since the partial order relation strictly
follows set inclusion, the lattice of properties looks as
the one drawn in
Fig.
\ref{figm4}.

\section{Concluding remarks}
\label{s5}

Let us finally state the
common features of our two approaches.

\paragraph{Operationalistic.}
Just as in quantum measurements,
the extrinsic view is assumed to be
inaccessible. All
we are allowed to do is to choose the input  to be applied and then
observe the output values.
For instance, in order to measure the electron spin, we are not allowed
to ``screw open'' the electron. Moreover, one cannot in general copy an
(arbitrary number of) identical electrons from the single electrons
taken from an  ensemble.

\paragraph{Mathematical.}
In both approaches, the collection of propositions is associated with a
partially
ordered set, whose elements are subsets of the set of states of the
automaton.

Since our two approaches are based on different empirical setups,
there are certain
features in which of our two approaches differ.

\paragraph{Automaton model.}
The normalized automata used here are appropriate for the logical
simulation of macrostatements. They are  assumed to have only two output
values.
With a different automaton model, to which the partition logic of
microstatements is applicable,
one obtains more
general structures of microstatements than the lattices $MOn$.
Examples of such automaton models are
 Moore
and Mealy type automata having more than two output values and thus
allowing for more than two elements in the partitions.
\paragraph{Operationalistic.}
\begin{itemize}
\item
To formulate the microstatements (to which partition logic is
applicable),
the experiment is assumed to be carried out with a
{\em single} automaton.
\item
Macrostatements refer to ensembles of automata.
\end{itemize}

\paragraph{Mathematical.}
\begin{itemize}
\item
In partition logic, the structure of the collection of properties is
always a
pasting of Boolean algebras. It may not even be a lattice
\cite[pp. 137-141]{svozil-93}.
\item
The structure of the collection of macroproperties is always a lattice
with orthocomplementation. It is not necessarily orthomodular
\cite{qwnhpa}.
\item
The collections of subsets associated with testable microstatements
and macrostatments do not in general include one another
but always overlap.
The partial ordering of these subsets can be induced
from both sorts of propositional structures.
For macrostatements, the partial order is exactly the set inclusion.
As has already been pointed out, for microstatements
the set inclusion implies the partial order
only if both statements are
simultaneously measurable (cf. section \ref{s3}).
\end{itemize}


\begin{thebibliography}{99}
 \bibitem{bauer}
Brauer, W. {\sl Automatentheorie} (Teubner, Stuttgart, 1984).
\bibitem{conway}
Conway, J.~H.: {\em Regular Algebras and Finite Machines},
(Clowes and Sons, London, Colchester, 1971).
\bibitem{finkelstein} Finkelstein D., and S.R.Finkelstein,
{\em Computational Complementarity}, International Journal of
Theoretical Physics, {\bf 22},
753, 1983.
\bibitem{grzap90} Grib A.A., and R.R. Zapatrin,{\em Automata Simulating
Quantum Logics}, International Journal of Theoretical Physics, {\bf 29},
113, 1990.
 \bibitem{grzap92} Grib A.A., R.R. Zapatrin, {\it
Macroscopic realizations of
quantum logics.}  Inter. J. Theor.Phys. {\bf 31} (1992), 1669--1687.
\bibitem{moore} Moore, E.F., {\em Gedankenexperiments on sequential
machines}, in {\em Automata Studies}, C.E.Shannon and J.McCarthy, eds.,
Princeton Univ. Press, Princeton (1956) 129-153.
\bibitem{svosh} Schaller M., K. Svozil, {\it Partition Logics of
Automata.} Il Nuovo Cimento {\bf 109 B} (1994) 167--176.
\bibitem{svosh2} Schaller M., K. Svozil,
{\it Automaton partition logic versus quantum logic.}
 Inter. J. Theor. Phys, {\it in print}.
\bibitem{svosh3} Schaller M., K. Svozil,
{\it Automaton logic.}
 Inter. J. Theor. Phys, {\it in print}.
\bibitem{svozil-93}
Svozil K.,  {\sl Randomness and Undecidability in Physics}
(World Scientific, Singapore, 1993).
\bibitem{qwnhpa} Zapatrin R.R., {\em Quantum Logics Without
Negation}, Helvetica Physica Acta, {\bf 67}, 188, 1995.
\end{thebibliography}
\end{document}